\documentclass[12pt]{iopart}
\usepackage[T1]{fontenc}					        	%vor inputenc
\usepackage[greek,ngerman,english]{babel}	        	%vor inputenc
\usepackage[utf8x]{inputenc}
\usepackage{lmodern}             		        		% Improved font rendering (esp. in PDFs)

\expandafter\let\csname equation*\endcsname\relax       % No Error while using mhchem with iopart class (before amsmath)
\expandafter\let\csname endequation*\endcsname\relax    % No Error while using mhchem with iopart class (before amsmath)
\usepackage{amsmath}

\usepackage{amsfonts}
\usepackage{amssymb}
\usepackage[]{graphicx}
\usepackage[bindingoffset=1cm,hmargin=2cm,vmargin=3cm]{geometry}
\usepackage{cite}
\usepackage{xcolor}							        %Farbiger Text \color{blue} oder \textcolor{red}{roter text}
\graphicspath{{graphics/}}

\usepackage{siunitx}                                %Nmmern mit SI Enheiten und geschütztem halb-Leerzeichen
\usepackage[version=4]{mhchem} 				        %Hübsche chemische Formeln \ce{H2O}
\usepackage{xspace} 							
\begin{document}

\newcommand{\MoS}{\ce{MoS2}\xspace}
\newcommand{\Al}{\ce{Al2O3}\xspace}
\newcommand{\Si}{\ce{SiO2}\xspace}
\newcommand{\SiO}{\ce{SiO2}/\ce{Si}\xspace}

\title{Growth of p-doped 2D-\ce{MoS2} on metal oxides from spatial atomic layer deposition}
\author{André Maas$^1$, Kissan Mistry$^2$, Stephan Sleziona$^1$, Abdullah H. Alshehri$^3$, Hatameh Asgarimoghaddam$^2$, Kevin Musselman$^2$, Marika Schleberger$^1$}
\address{$^1$Universität Duisburg-Essen, Fakultät für Physik and CENIDE, Germany}
\address{$^2$Department of Mechanical and Mechatronics Engineering, University of Waterloo, Canada}
\address{$^3$Department of Mechanical Engineering, Prince Sattam bin Abdul Aziz University, Alkharj 11942, Saudi Arabia }

\begin{abstract}
In this letter we report on the synthesis of monolayers of \MoS via chemical vapor deposition directly on thin films of \Al grown by spatial atomic layer deposition. The synthesized monolayers are characterized by atomic force microscopy as well as confocal Raman and photoluminescence spectroscopies. Our data reveals that the morphology and properties of the 2D material differ strongly depending on its position on the substrate. Close to the material source, we find individual flakes with an edge length of several hundred microns exhibiting a tensile strain of \SI{0.3}{\percent}, n-doping on the order of $n_e$=\SI{0.2e13}{cm^{-2}}, and a dominant trion contribution to the photoluminescence signal. In contrast to this, we identify a mm-sized region downstream, that is made up from densely packed, small \MoS crystallites with an edge length of several microns down to the nanometer regime and a coverage of more than \SI{70}{\percent}. This nano-crystalline layer shows a significantly reduced  strain of only \SI{<0.02}{\percent}, photoluminescence emission at an energy of \SI{1.86}{eV} with a reduced trion contribution, and appears to be p-doped with a carrier density of $n_h$=\SI{0.1e13}{cm^{-2}}. The unusual p-type doping achieved here in a standard CVD process without substitutional doping, post-processing, or the use of additional chemicals may prove useful for applications.
\end{abstract}
\maketitle

\section{Introduction}
In recent years there has been an increasing interest in 2D materials for high performance electronic and optoelectronic applications \cite{Chen.2016,Yu.2013,Zhang.2014}, including flexible devices \cite{Yu.2014,Amani.2015,Musselman.2016} and field-effect transistors \cite{Bergeron.2017}. An important class of 2D materials is transition metal dichalcogenides (TMDC) with the chemical formula MX$_2$. The hexagonal structure is formed by a sheet of transition metal (M) atoms covalently bound in between two layers of chalcogen (X) atoms. Some TMDCs like molybdenum disulfide (\MoS) have semiconductor properties like an indirect band gap in the near infrared. When their spatial extension is confined in one dimension this band gap becomes a direct band gap in the visible spectrum \cite{Mak.2010,Kuc.2011,Splendiani.2010}, making these materials even more attractive for (flexible) optoelectronics. This however requires the fabrication of heterostructures, i.e.,~the combination of the TMDC with at least one dielectric material. The current approach of layering \MoS on top of high-$\kappa$ dielectrics utilizes growth via chemical vapour deposition (CVD) on selected substrates, typically \Si, and subsequent transfer onto the final dielectric materials. This process suffers from various limitations including the loss of structural integrity of the \MoS layer, no control over the doping, and contamination through residuals from the transfer process \cite{Pollmann.2020,Pollmann.2020b}. Growing the \MoS directly on top of the dielectric could overcome these severe limitations \cite{Bergeron.2017}.

While this approach seems straightforward, it is still a challenge to obtain a continuous layer of \MoS using a common fabrication method such as CVD. The resulting monolayers are typically highly crystalline, but their lateral dimensions rarely exceed a few hundred microns \cite{Bergeron.2017,Zhou.2018}. Also, the large-scale fabrication of tailored dielectrics poses a major problem, in particular if the flexibility of the 2D material is to be maintained, requiring ultrathin dielectric films. To address both problems it has recently been tried to combine CVD growth of \MoS with the synthesis of the dielectric material via atomic layer deposition \cite{Bergeron.2017}. The resulting flakes were on average rather small (edge length of a few microns), could be processed into a working field-effect device and showed a PL signal at room temperature (RT). Atomic layer deposition (ALD) is already in use worldwide for the fabrication of microelectronics and can deposit nanolaminated stacks of films \cite{Kattelus.1993,Kukli.1997,Kukli.1996}. However, one of the biggest drawbacks of this technique is its low deposition rate and therefore it may not always be commercially viable considering the price and its limited throughput \cite{Liddle.2016}.
 
We therefore suggest to use the newly developed process of atmospheric-pressure spatial atomic layer deposition (AP-SALD) to overcome this drawback. This technique separates the precursor gases not temporally but spatially, making it a faster and more cost-effective way of creating a thin layer with sufficient control over the layer thickness \cite{Musselman.2016,Mistry.2020,Poodt.2010}. AP-SALD has already been successfully utilized for the growth of uniform films for different applications including solar cells \cite{Musselman.2014,Ievskaya.2016}, light emitting diodes \cite{Hoye.2015b,Hoye.2015,Di.2017}, and quantum tunneling diodes \cite{Alshehri.2019}. Here, we report on the first results of growing a 2D material on a dielectric film deposited by AP-SALD. Monolayer MoS$_2$ is grown via CVD on thin Al$_2$O$_3$ films deposited by AP-SALD. By a detailed analysis we show that the combination of those two scalable approaches may be used to fabricate mm-sized samples largely covered with nano-crystalline p-doped 2D-\MoS. %\textcolor{blue}{An dem Schlusssatz muss man ggf. noch arbeiten.}
 
\section{Results and Discussion}
\begin{figure}[htb]
    \centering
    \includegraphics[width=0.8\textwidth]{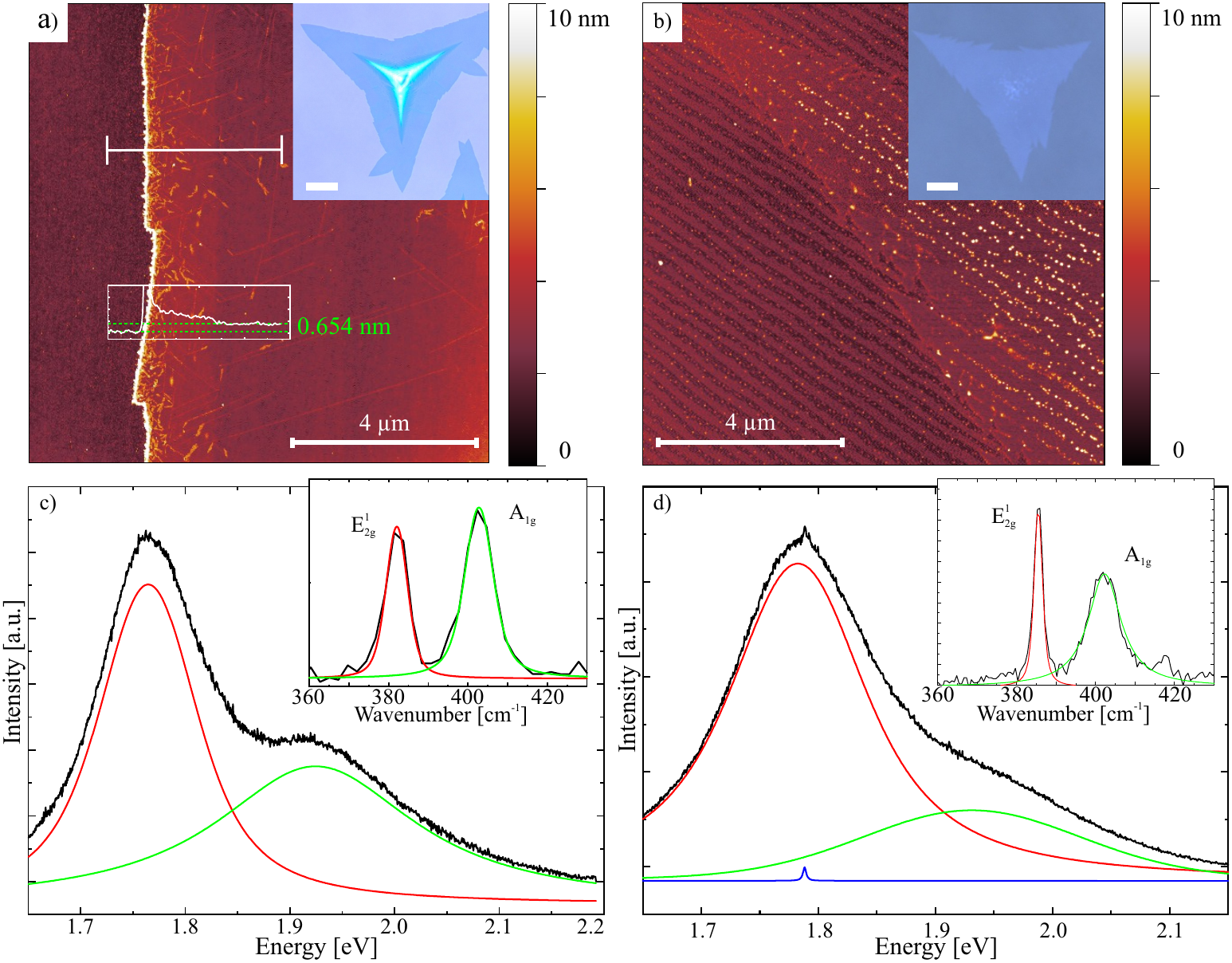}
    \caption{AFM/optical images and PL/Raman spectra of \MoS monolayers grown via CVD on the two most common substrates. AFM of \MoS on \Si (a) and on sapphire (b), and the corresponding optical images (inset). The scale bar is \SI{10}{µm}. The height of the layer was measured to be \SI{0.654}{nm} (a, line, white), which is consistent with the nominal single layer height of \MoS. (b) clearly shows the atomic terraces of the single crystalline substrate. The Pl spectra from both samples show a distinct RT photoluminescence at 1.76 eV (\Si) and 1.8 eV (\Al), respectively, that is characteristic for the \MoS monolayer. The Raman spectra show the fingerprint A$_{1g}$ and E$^1_{2g}$ vibrational modes.}
    \label{fig:ORPC}
\end{figure}

The growth of TMDCs via CVD is typically dominated by the material supply. This in turn is governed by the material sources, the TMDC itself, and the substrate. The most common substrate is thermally grown \Si on top of a Si wafer, which allows the fabrication of field-effect transistors (FET) based on the as grown-material without an additional transfer step. This may however not be an ideal substrate for the synthesis of large MoS$_2$ samples, as thermally grown \Si is amorphous and and exhibits a typical roughness (root mean square) of $z_{RMS}=0.5$~nm. Thus, if the fabrication on a FET is not the final goal, better suited substrates can be used which may facilitate large-scale growth. For \MoS, epi-polished, single crystals of \Al have proven to be an almost ideal substrate with atomic scale roughness and a lattice constant of $a_{Al2O3}=4.8$~\AA. The lattice is commensurable to the MoS$_2$ lattice ($a_{MoS2}=3.2$~\AA \cite{N.Wakabayashi1975}) and mm-sized, epitaxial monolayers could be grown via CVD \cite{Dumcenco.2015}.
As reference systems for our study, we have therefore grown MoS$_2$ via CVD on amorphous \Si (see Fig.~\ref{fig:ORPC}a) and on crystalline \Al (see Fig.~\ref{fig:ORPC}b; for details see method section). On both substrates, individual flakes with the typical star-like triangular form have grown. On average the flakes on \Si are somewhat smaller than the ones on \Al. Both samples exhibit the typical in- and out-of-plane Raman modes E$^1_{2g}$ and A$_{1g}$ at wavenumbers 382~cm$^{−1}$ and 403~cm$^{−1}$ for the \Si substrate, and 385~cm$^{−1}$ and 402~cm$^{−1}$ for the sapphire substrate, respectively, as well as a strong photoluminescence signal at an energy around $E=1.8$~eV, the latter confirming the materials´ 2D nature. From the position of the Raman modes we deduced the strain and doping of the grown flakes following the approaches presented in Refs. \cite{Pollmann.2018,Panasci.2021,Schiliro.2021}. We find that \MoS on \Si exhibits a tensile strain of \SI{0.45}{\percent} and is highly n-doped with $n_e$=\SI{1.0e13}{cm^{-2}}, and on sapphire it shows a compressive strain  of \SI{-0.3}{\percent} and is similarly n-doped with $n_e$=\SI{1.2e13}{cm^{-2}}. 

\begin{figure}[htb]
    \centering
    \includegraphics[width=0.9\textwidth]{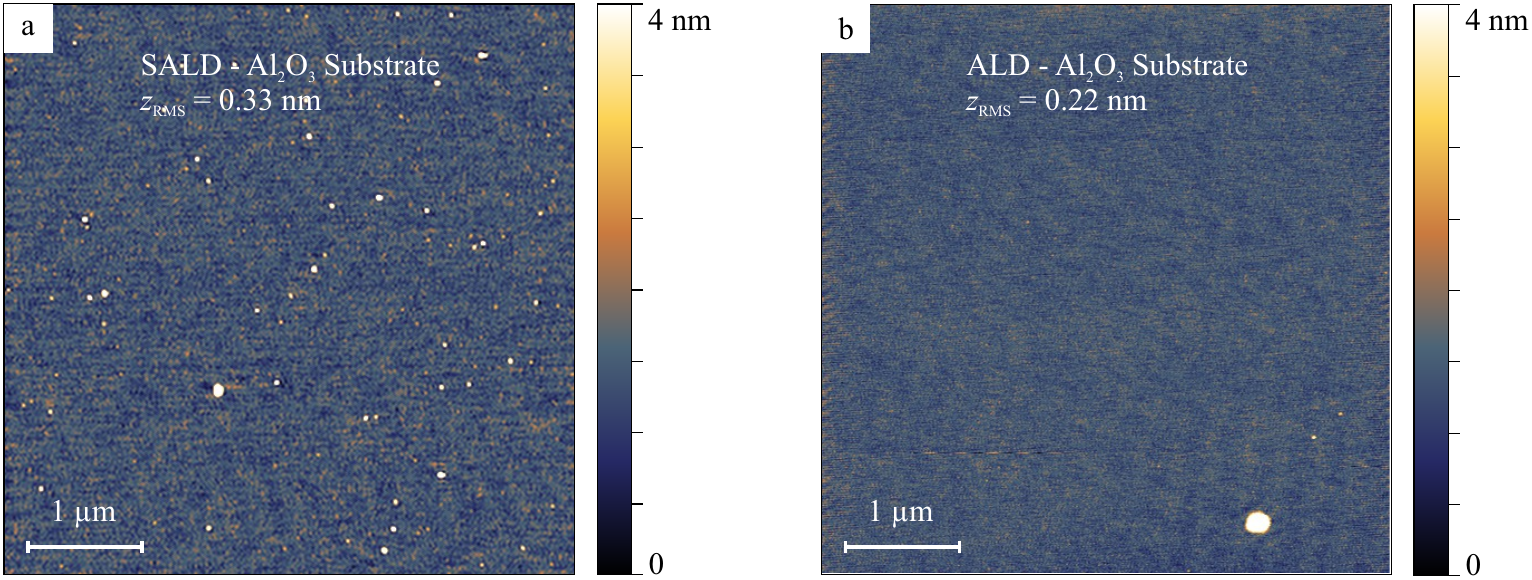}
    \caption{AFM images of the \Al substrate deposited by AP-SALD (a), conventional ALD (b). In both cases a uniform, continuous layer is formed. The statistical analysis of the height data yields a roughness of about $z_{RMS} = 0.33$~nm for the AP-SALD layer and $z_{RMS} =$0.22~nm for the ALD layer.}
    \label{fig:ald_sald}
\end{figure}

From Fig.~\ref{fig:ORPC} it is obvious that in principle the amorphous structure of \Si does not prevent the synthesis of large monolayer flakes and thus it seems promising to investigate the possibility to grow TMDCs on amorphous \Al. To this end, a \ce{Si} wafer with a \SI{385}{nm} \ce{SiO2} layer on top was used as a base substrate. Then, \SI{20}{nm} of \Al were deposited on top of the wafer via AP-SALD (for details, see method section), resulting in a seemingly continuous layer as shown in Fig.~\ref{fig:ald_sald}a. The thickness can be chosen freely within the limits of the method and in particular it can be chosen to be sufficiently thin to enable flexible devices. For the purpose of this study we have chosen 20~nm which should be thick enough to prevent the underlying \ce{SiO2} layer from influencing the \MoS growth, e.g.~by interface charges \cite{Schiliro.2021}. The \Al layers deposited in this way have an average roughness of $z_{RMS}=\SI{0.33}{nm}$, which is slightly larger than for a layer grown by conventional ALD ($z_{RMS}=\SI{0.22}{nm}$, shown for comparison in Fig.~\ref{fig:ald_sald}b), but comparable to the roughness of a clean, amorphous \ce{SiO2} surface. 

%\textcolor{blue}{Dies könnte ggf. auch in eine SI; dort dann auch SiO2.}

\begin{figure}[htb]
    %\centering
    \includegraphics[width=1.0\textwidth]{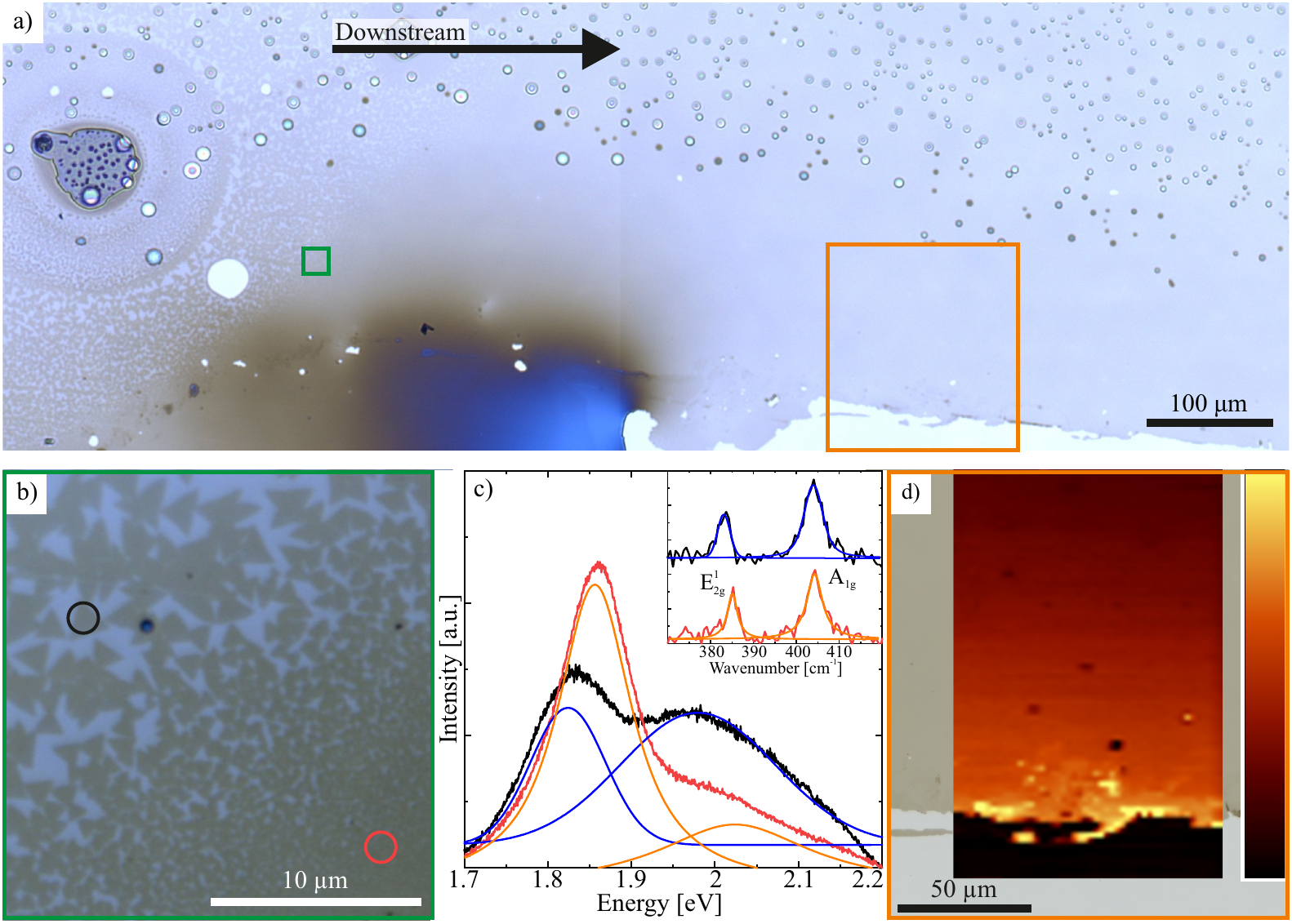}
    \caption{Optical images, exemplary Raman, PL spectra and mapping of \MoS monolayers grown via CVD on the AP-SALD-grown \Al layer. Optical image (a) of the substrate fabricated by AP-SALD after the \MoS process. Downstream direction is marked with arrow. Separated flakes (upstream) are visible as well as a continuous layer of flakes (downstream). The transition area in between (green square) is shown in (b). The individual flakes have a triangular structure; Raman and PL spectra shown in (c) are taken are from the areas marked by the circles. The PL peaks for the nano-crystallites (b, red) have more intensity and are more narrow than for the individual flakes (b, black). The Raman spectra show the typical \MoS modes for both (c, inset). The E$^{\text{1}}_{\text{2g}}$ and A$_{1g}$ modes where measured at \SI{383}{cm^{-1}} and \SI{404}{cm^{-1}} for the individual flakes and at \SI{385}{cm^{-1}} and \SI{404}{cm^{-1}} for the nano-crystallites. A PL mapping (d) shows the intensity at \SI{1.86}{eV} and that the nano-crystalline region expands over several hundred microns. In the lower part, where no PL activity is measured, no a-\Al was deposited and no \MoS growth was detected.} 
    \label{fig:ORP}
\end{figure}

\begin{figure}[htb]
    \centering
    \includegraphics[width=1.0\textwidth]{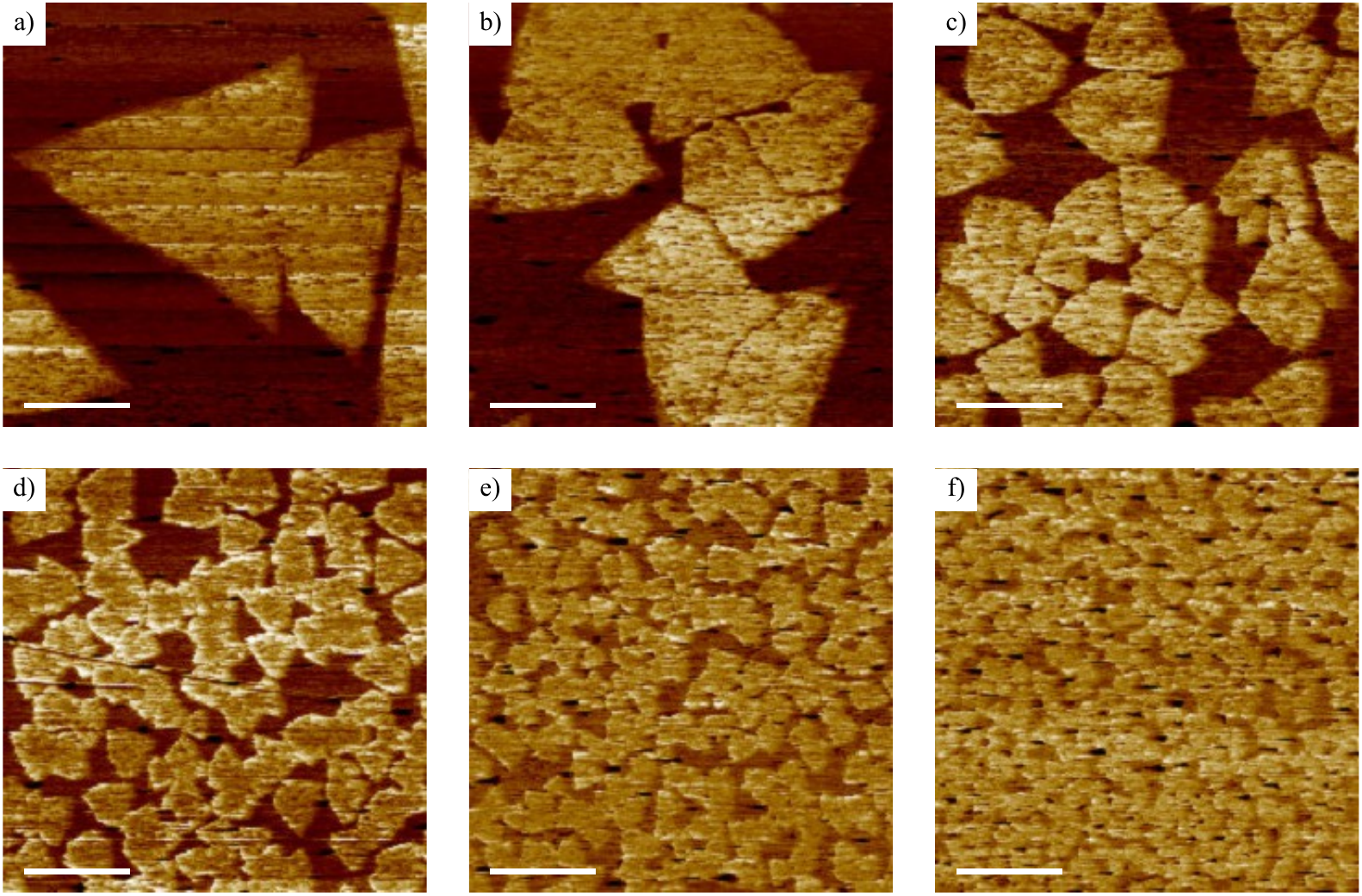}
    \caption{Peak-force-microscopy images taken in different regions of the sample starting from micron-sized individual flakes (a-c) in the upstream region to the nano-crystallites typically found in the downstream region (d-f). The filling factor increases from \SI{49}{\percent} (a) to \SI{73}{\percent} (f). The scale bar is \SI{1}{µm}.}
    \label{fig:afm_comp}
\end{figure}

The AP-SALD substrate was then used in a standard CVD process to deposit \MoS (for details, see method section). In Fig.~\ref{fig:ORP}a an optical image from the as-grown \MoS is shown. At the top left side of the sample, close to the material source, the \MoS has grown as individual flakes that have the typical triangular shape with a size of up to \SI{100}{µm}. Moving to the bottom right, i.e., further downstream, one can see that the density of the flakes increases while their size decreases. This clearly shows that the amorphous AP-SALD substrate allows for very different flake morphologies, depending on the temperature and material supply. In particular, there is a large region, extending roughly from the middle of the first third of the substrate (position of the green square) to its lower left right edge, where the flake size appears to be too small to be resolved by optical microscopy and a seemingly continuous film has been synthesized. Note, that this region is on the order of several square millimetres. To the best of our knowledge, this has never been achieved with an amorphous \SiO substrate and furthermore did also not occur on ALD-grown \Al substrates (see SI). Further analysis with AFM reveals the true morphology in this region. As shown in Fig. \ref{fig:afm_comp}, the seemingly uniform film is made up from densely packed nano-crystallites with an average height of $\simeq 7$~\AA. The edge length of the individual flakes decreases from several microns (Fig. \ref{fig:afm_comp} a-c) down to <\SI{100}{nm} (Fig. \ref{fig:afm_comp} d-f) within a span of \SI{120}{µm}. At the same time the filling factor increases from \SI{35}{\percent} for the individual flakes to about \SI{75}{\percent} for the nano-crystallites. 

To further verify that this nano-crystalline film region is made up from \MoS monolayer flakes, Raman and PL spectroscopies were performed. For easy comparison, we chose the transition region (green square in Fig.~\ref{fig:ORP}) and start upstream with an analysis of the individual flakes (Fig.~\ref{fig:ORP}b, black circle), where the Raman E$^{\text{1}}_{\text{2g}}$ and A$_{1g}$ modes are found at $\SI{383}{cm^{-1}}$ and $\SI{404}{cm^{-1}}$, i.e., exhibiting a difference of $\SI{21}{cm^{-1}}$ (Fig.~\ref{fig:ORP}c, black curve in the inset). The difference in wave numbers confirms that these flakes are single layer \MoS \cite{Lee.2010}, and this is corroborated by a moderate PL at \SI{1.8}{e.V.} (Fig.~\ref{fig:ORP}c, black curve). The PL signal shows a strong trion contribution at $\simeq$~\SI{2.0}{eV}, as shown by the blue fits in Fig.~\ref{fig:ORP}c, that points towards a significant density of excess charges. From the Raman mode positions we again infer that the \MoS here is single layer \cite{Lee.2010}, n-doped with $n_e$=\SI{0.2e13}{cm^{-2}}, and experiences a tensile strain of \SI{0.3}{\percent} \cite{Pollmann.2018,Panasci.2021,Schiliro.2021}. The strain is comparable to that observed on the \Si substrate (see Fig.~\ref{fig:ORPC}), but the doping is reduced quite drastically when compared to both \Si and sapphire as substrates.  

Next, we will analyze the nano-crystalline region located further downstream (Fig.~\ref{fig:ORP}b, red circle). Note, that the overall intensity of the \MoS Raman modes (Fig~\ref{fig:ORP}c, inset) is in general very low, compared to the \ce{Si} signal at $\SI{520}{cm^{-1}}$, which is probably due to the different optical properties of the AP-SALD substrate \cite{Zhang.2015}. The different intensity ratio of the in-plane E$^{\text{1}}_{\text{2g}}$ mode and the A$_{1g}$ mode maybe due to the different polarization dependence of the two modes \cite{Lee.2010}. 

Despite the comparatively poor signal-to-noise ratio we were able to perform a quantitative analysis. We find that both in- and out-of-plane mode have shifted in comparison to the modes observed from individual flakes (see red and black curves in the inset of Fig.~\ref{fig:ORP}c), exhibiting an even smaller difference of 20~cm$^{-1}$. The E$^{\text{1}}_{\text{2g}}$ mode shifts to \SI{385}{cm^{-1}} which shows that the strain of the material is now negligible with \SI{<0.02}{\percent} and thus comparable to the reference value of freestanding \MoS \cite{Lloyd.2016,Panasci.2021}. This relaxation is probably connected to the reduced size of the flakes. The A$_{1g}$ mode red-shifts to \SI{405}{cm^{-1}} and therefore the material here is p-doped with $n_h$=\SI{0,1e13}{cm^{-2}} \cite{Panasci.2021,Xu.2017}. The observed change in doping is in striking contrast to values reported in literature so far, where \MoS usually shows n-type doping on \Al and SiO$_2$ \cite{Panasci.2021,Mak.2013,Radisavljevic.2011}, while p-type doping has been observed so far only on metallic substrates, as for example, \ce{Au} \cite{Panasci.2021,Pollmann.2021}. P-type doping of \MoS can be achieved in different ways \cite{Xin.2020,Xu.2017}, e.g., also by adding appropriate reactants in the CVD process but the resulting carrier density determined in a field-effect transistor was an order of magnitude lower \cite{Lee.2019}. 

Interestingly, in comparison, the PL in the nano-crystalline region has a significantly higher intensity at an energy of about \SI{1.85}{eV} (Fig.~\ref{fig:ORP}c, red curve). This considerable PL signal again confirms that the nano-crystallites consist of single layer \MoS \cite{Kuc.2011}. The trion contribution is strongly reduced  to $I_T/I_E=0.78$ ($I_T/I_E=0.54$ at 77 K, see SI) as derived from the orange fits in Fig.~\ref{fig:ORP}c. In comparison, the large flakes exhibit a ratio of $I_T/I_E=2.00$. The reduced trion contribution indicates that less excess charge carriers are available and mirrors the red-shift of the Raman modes associated with p-type doping in the nano-crystalline region. Low-temperature PL measurements (see SI) show that the peaks barely shift (around \SI{20}{meV}) which is to be expected for a non-strained monolayer. Finally, a PL mapping of a large sample region (Fig.~\ref{fig:ORP}d) shows that indeed the whole nano-crystalline film is PL-active. 

\section{Conclusion}
We have synthesized monolayers of \MoS via CVD on thin films of \Al grown via AP-SALD. The resulting flakes show a large variation in properties depending on their position on the substrate. While this effect is in principle well-known from all CVD processes \cite{Kumar.2018}, we observe a strong and quite unusual difference in morphology, and related to this, unexpected changes of vibrational and optoelectronic properties. In particular, while the larger, individual flakes show only weak, trion-dominated PL and appear n-doped and strained, we identified a region with a very high (up to \SI{73}{\percent}) coverage of nano-crystalline \MoS with negligible strain, p-type doping and a significant PL emission dominated by the A exciton. Whether this unusual doping can be maintained upon further processing remains to be seen. The unique morphology of \MoS monolayer nano-crystallites with a high density of edges combined with a high coverage appears exclusively on the AP-SALD-grown substrates and could be advantageous for, e.g., catalysis \cite{Madau.2018}, sensing \cite{Cho.2015} or electrochemical \cite{Li.2016} applications. In terms of (flexible) devices, it would be beneficial to further increase the coverage up to a completely continuous, closed film. One way to achieve this could be a second processing step, exploiting the fact that the growth is dominated by the transport of material across the flake area (and not along the edges) \cite{Pollmann.20.01.2022}, which could lead to a preferential growth in-between the pre-existing flakes. An alternative way could be to combine the AP-SALD-grown substrate with another variant of CVD, i.e, metal organic CVD which is known to produce uniform and closed layers of \MoS composed of many small crystallites \cite{Kang.2015}.

\section{Methods}

\subsection{Spatial Atomic Layer Deposition}
AP-SALD is a modification of the conventional ALD approach, in which different precursor exposure steps of conventional ALD are not temporally but spatially separated. In this work, a custom-built, linear “zone-separated” AP-SALD system was used. This arrangement works with a fixed reactor head through which the individual precursor gases pass in different channels. The gases exit in parallel streams at different locations on the bottom of the head creating precursor filled zones. The individual zones are separated by an inert nitrogen gas. Furthermore, in between the individual gas channels, there are exhaust channels through which the gases are removed using a vacuum pump. 
The substrate was placed on a heated substrate stage approximately \SI{100}{µm} below the reactor head and oscillated back and forth at \SI{50}{mm/s} to deposit a film with the desired \SI{20}{nm} thickness (\SI{80}{cycles}). By eliminating the vacuum chamber and purge steps, the throughput is higher and the process more cost effective than that of a conventional ALD system. 

Trimethylaluminum (TMA, Strem Chemicals) and distilled water were used as the precursors. To deliver the TMA to the reactor head, nitrogen was bubbled through the TMA at \SI{38}{sccm} and combined with a carrier nitrogen flow of \SI{87}{sccm}. Similarly, the distilled water was bubbled with a \SI{125}{sccm} nitrogen flow and combined with a carrier nitrogen flow of \SI{125}{sccm}. A flow of \SI{2000}{sccm} was delivered to 6 inert nitrogen gas channels separating the precursor channels. The heated substrate stage was held at \SI{250}{\degreeCelsius} throughout the deposition.

\subsection{Atomic Layer Deposition}
In this process, three precursors were employed: Trimethylaluminum (TMA), \ce{O2} plasma and Argon (Ar) as inert gas. \SI{50}{cycles} were used to deposit \SI{5}{nm} thick \Al film. The film was deposited at \SI{100}{\degreeCelsius} with a pressure of \SI{1.3e-4}{Pa}. The reactor chamber was first purged with Ar for \SI{3}{min} to stabilize the chamber temperature and pressure. Ar was bubbled through TMA to dose the substrate for \SI{20}{ms} at a pressure of \SI{2}{Pa}, and then the precursor line was purged with Ar for \SI{1}{s} at a pressure of \SI{2}{Pa}. After that, a flow of \ce{O2} was stabilized for \SI{500}{ms} before generating a plasma by RF with a power of \SI{300}{W} at a pressure of \SI{2}{Pa}. The substrate was then exposed to the \ce{O2} plasma for \SI{2}{s} and followed by purging for \SI{1}{s}.

\subsection{Chemical Vapor Deposition}
The custom-made synthesis protocol is based on the system by Lee et al.\cite{Lee.2012} and is used to fabricate crystalline \MoS monolayers on various substrates\cite{Pollmann.2018}. The \Si substrate was a p-doped silicon wafer in (100) orientation with a \SI{285}{nm} oxide thickness from \textit{Graphene Supermarket}. The Sapphire substrate was a c-plane (0001), epi-polished monocrystalline \Al wafer with a \SI{0.2}{\degree} offcut from \textit{Roditi}. It was annealed at \SI{1000}{\degreeCelsius} for \SI{60}{min} prior to synthesis.

The molybdenum source was ammonium heptamolybdate (AHM, Sigma Aldrich), an inorganic ammonium salt soluble in water. The solution with \SI{200}{\gram\per\liter} was deposited onto the substrate in a single droplet of \SI{1}{\mm} in diameter at the upstream edge of the substrate. This substrate including the precursor material was heated to \SI{300}{\degreeCelsius} for \SI{30}{\min} to form \ce{MoO3} \cite{Hanafi.1981}. In a \SI{1}{Vol\%} solution the seeding promoter, cholic acid sodium salt (Sigma Aldrich), was spin coated onto the substrate containing the \ce{MoO3} and put into a crucible. A second crucible was filled with \SI{40}{\milli\gram} sulfur (S powder, Sigma Aldrich, \SI{99.98}{\%}). The chemical vapor deposition took place in a tube furnace (ThermConcept, ROK 70/750/12-3z) with three thermally separated heating zones. The crucible containing the substrates was placed into the second heating zone (downstream) and the crucible containing the sulfur was placed into the first heating zone (upstream). The tube was purged with inert argon gas. For the synthesis, the Ar flow rate was set to \SI{500}{sccm}. The downstream zone was heated to \SI{750}{\degreeCelsius} in \SI{10}{\minute} and held at that temperature for \SI{20}{\minute}. The upstream zone is heated to \SI{150}{\degreeCelsius} in \SI{10}{\minute} and held at that temperature for \SI{20}{\minute}. The process is ended by rapidly cooling down the samples.

\subsection{Atomic force microscopy}
Atomic force microscopy images were taken with a \textit{Veeco Dimension 3100 AFM} in tapping-mode using Nanosensors NCHR-50 tips. The peak-force-microscopy measurements were performed with a \textit{Bruker Dimension Icon} in the PeakForce Tapping Mode using ScanAsyst-Air tips. The AFM results were analyzed and visualized by \textit{Gwyddion 2.60}. The filling factors were evaluated by increasing the contrast between flakes and substrate until a black and white image was obtained and the counting the respective pixels.

\subsection{Raman and Photoluminescence spectroscopy}
Raman and PL spectroscopy were performed with a \textit{WiTec alpha300 RA} confocal Raman spectrometer. All measurements were performed with \SI{532}{nm} excitation wavelength and an output power of \SI{4}{mW}. For Raman measurements a \SI{1800}{cm^{-1}} grid, and for PL measurements a \SI{300}{cm^{-1}} grid was used, respectively. 
%Low temperature measurements were performed inside a \textit{Linkam Stage THMS350EV} which is compatible with the used Raman system.

\section*{Acknowledgment}
We thank ICAN (S.~Franzka) for conducting the peak-force-microscopy measurements and acknowledge financial support from the Federal Ministry of Education and Research (BMBF, project 05K19PG1), the German Research Foundation (DFG, SCHL 384/20-1, project number 406129719) and from CENIDE. K.P.M. acknowledges funding from NSERC Discovery (RGPIN-2017-04212, RGPAS-2017-507977), Canada Foundation For Innovation John R. Evans Leaders Fund (Project 35552), and Ontario Ministry of Research, Innovation and Science ORF-Small Infrastructure (Project 35552). The University of Waterloo’s QNFCF facility was used for this work.

\newpage

%******		Bibliography 	***********************************
%\addcontentsline{toc}{chapter}{References}			%References in table of content 						
\bibliography{Citavi} 								%1x BibTeX (F11) , 2x Schnelles Üb. (F1)
\bibliographystyle{ieeetr}

%\nocitenames 										%enable to show not cited names
%\nocite{*}											%enables the marking not cited/ labled names and equasions
%**************************************************************

\end{document}